\documentclass{elsart4-1}

\usepackage{amsmath, amssymb}

\usepackage[english,final]{LLNdef}
\usepackage{journames}
\usepackage[final]{modifications}
\usepackage{cite}
\usepackage{citesort}

\ifx\pdfoutput\undefined\usepackage{graphicx}\else\usepackage{graphicx}\fi
\graphicspath{{img/150dpi/}}

\usepackage[english,francais]{babel}

\newtheorem{e-proposition}[theorem]{Proposition}

\newtheorem{e-definition}[theorem]{Definition\rm}

\def\og{\leavevmode\raise.3ex\hbox{$\scriptscriptstyle\langle\!\langle$~}}
\def\fg{\leavevmode\raise.3ex\hbox{~$\!\scriptscriptstyle\,\rangle\!\rangle$}}

\begin{document}

\begin{frontmatter}


\selectlanguage{english}
\title{Epitaxial self-organization: from surfaces to magnetic materials}

\vspace{-2.6cm}

\selectlanguage{francais}
\title{Auto-organisation {\'e}pitaxiale: des surfaces aux mat{\'e}riaux magn{\'e}tiques}


\selectlanguage{english}
\author[of]{Olivier Fruchart}
\ead{Olivier.Fruchart@grenoble.cnrs.fr}

\address[of]{Laboratoire Louis N{\'e}el (CNRS) -- 25, avenue des Martyrs -- BP166 -- F-38042 Grenoble Cedex 9 -- France}

\begin{abstract}
Self-organization of magnetic materials is an emerging and active field. An overview of the use of
self-organization for magnetic purposes is given, with a view to illustrate aspects that cannot be
covered by lithography. A first set of issues concerns the quantitative study of low-dimensional
magnetic phenomena~(1D and 0D). Such effects also occur in microstructured and
lithographically-patterned materials but cannot be studied in these because of the complexity of
such materials. This includes magnetic ordering, magnetic anisotropy and superparamagnetism. A
second set of issues concerns the possibility to directly use self-organization in devices. Two
sets of examples are given: first, how superparamagnetism can be fought by fabricating thick
self-organized structures, and second, what new or improved functionalities can be expected from
self-organized magnetic systems, like the tailoring of magnetic anisotropy or controlled
dispersion of properties. {\it To cite this article: O. Fruchart, C. R. Physique 6 (1) (2005).}

\vskip 0.5\baselineskip

\selectlanguage{francais}
\noindent{\bf R{\'e}sum{\'e}}%
\vskip 0.5\baselineskip%
\noindent%
Alors que l'auto-organisation est un domaine maintenant consacr{\'e} pour les semi-conducteurs, il est
en {\'e}mergence pour les mat{\'e}riaux magn{\'e}tiques, avec une activit{\'e} soutenue les cinq derni{\`e}res ann{\'e}es.
Un panorama des contributions de l'auto-organisation au magn{\'e}tisme est propos{\'e} ici, avec pour but
de montrer les possibilit{\'e}s nouvelles offertes, notamment par rapport {\`a} la lithographie. Une
premi{\`e}re cat{\'e}gorie d'{\'e}tudes concerne la mesure et la compr{\'e}hension de ph{\'e}nom{\`e}nes magn{\'e}tiques en
basse dimensionnalit{\'e}, qui existent dans les mat{\'e}riaux applicatifs mais ne peuvent y {\^e}tre {\'e}tudi{\'e}s
quantitativement du fait de leur complexit{\'e}~: mise en ordre magn{\'e}tique, anisotropie magn{\'e}tique,
superparamagn{\'e}tisme. Une seconde cat{\'e}gorie concerne la perspective de l'utilisation directe de
syst{\`e}mes auto-organis{\'e}s. Des exemples sont donn{\'e}s pour combattre le superparamagn{\'e}tisme en
fabriquant des structures auto-organis{\'e}es {\'e}paisses, ou {\'e}tablir des fonctionnalit{\'e}s nouvelles,
notamment le contr{\^o}le de l'anisotropie et de la dispersion de propri{\'e}t{\'e}s. {\it Pour citer cet
article~: O. Fruchart, C. R. Physique 6 (1) (2005).}

\vskip 0.5\baselineskip%
\noindent{\small{\it Mots-cl{\'e}s~:} Auto-organisation; auto-assemblage; magn{\'e}tisme; anisotropie magn{\'e}tique; micromagn{\'e}tisme; superparamagn{\'e}tisme}%
}\end{abstract}
\end{frontmatter}

\selectlanguage{english}
\section{Introduction}

It was the semiconductor community who first drew its attention on deposition processes yielding
spontaneously nanostructures at surfaces. This alternative approach to lithography, a so-called
\emph{bottom-up} approach, may be called either self-assembly~(\SA) or
self-organization~(\SO)\cite{bib-CRphys-SOSA}. This topic was initiated in the mid-eighties with
the prospect of fabricating high-efficiency lasers with quantum dots\cite{bib-ARA82,bib-ASA86} and
is still a hot topic today\cite{bib-CRphys-Mariette}, with however modified
prospects\cite{bib-REI03,bib-BER03b,bib-MIC03}. The first demonstrations of \SA\ and \SO\ for
magnetic materials date back to the early nineties. The topic has become very active only in the
past few years, motivating the present overview of the contribution of \SA\ and \SO\ to the
advancement of magnetism.

Mostly single-element metallic systems have been demonstrated for magnetic materials, with however
recent reports on oxydes\cite{bib-LUD04,bib-ZHE04,bib-VAS03}, metallic alloys\cite{bib-ALB01} and
metals on molecular templates\cite{bib-MA04}. Some publications aimed at showing that these growth
phenomena occur for magnetic material, \eg, \SO Ni, Fe and Co dots on
Au(111)\cite{bib-CHA91,bib-CUL99,bib-VOI91}, \SO mono-atomic Fe stripes on
Au(788)\cite{bib-SHI04}, \SO\ Fe and Co dots on reconstructed N-Cu(001)\cite{bib-PAR97,bib-SIL00},
Ni stripes on $\mathrm{Cu}(110)-(2\sqrt{2}\times\sqrt{2})R45\deg-\mathrm{O}$\cite{bib-FUJ98}, Fe
dots and wires on $\mathrm{H}_2\mathrm{O}/\mathrm{Si}(100)(2\times n)$\cite{bib-KID99}, \SA\ Fe
dots on flat NaCl\cite{bib-GAI02}. Some other publications reported magnetic measurements in these
structures, however with no specific purpose, \eg, for \SO\ Co dots on Au(111)\cite{bib-TAK97},
\SO\ Fe dots and wires on facetted NaCl\cite{bib-SUG97}, \SA\ TM-RE dots on
Nb(110)\cite{bib-MOU00}. Neither these growth nor magnetic studies will be reviewed here. As the
field is becoming riper specific uses are being sought for \SA and \SO magnetic systems, with the
question whether they might be useful for fundamental science and/or for applications. In other
words, the question is: can one reconcile surface-science growth and magnetic investigations, with
materials' knowledge and applications?

The use of magnetic \SA\ and \SO\ can be classified into two categories. In the first category
such structures are used to gain information about fundamental phenomena that occur in materials
and systems that may be of interest for applications, but cannot be understood directly, because
they are too complex~(owing to microstructure, defects, size, \ldots). In this case \SA-\SO\
systems are used as objects of very high quality to serve as model systems for analyzing
fundamental issues of magnetism, preferably to lithography or microstructured materials. The major
investigated issues are magnetic order and thermal excitations in reduced
dimension\bracketsecref{sec-magneticOrder}, the crossover from bulk towards single atoms for spin,
orbital momentum and magnetic anisotropy energy~(\MAE) \bracketsecref{sec-anisotropy}, and finally
micromagnetism\bracketsecref{sec-micromagnetism}. These studied are of applied interest, because
devices require the use of ever smaller nanostructures, whose properties must be understood and
ultimately tailored. The second category consists in investigating whether \SO\ and \SA\ systems
might be used directly for applications. One fundamental obstacle forbidding this is the loss of
most magnetic functionalities at room temperature for very small systems due to thermal
excitations. This motivated the development of growth processes that replicate vertically
initially flat \SO\ structures to increase their volume with no compromise on their lateral
size\bracketsecref{sec-thick}. Also, examples are given were \SA-\SO\ can be used to achieve
materials with specific magnetic properties\bracketsecref{sec-material}.

For these two categories \SA\ or \SO\ systems might be used, although \SO\ generally receives
more attention. For a broad public the order and its fascinating beauty certainly account for
this. From a scientific point of view, an organized system is associated with a low dispersion
of size and shape, and thus of physical properties. This benefits both to fundamental
investigations because one can measure large assemblies and consider macroscopic measurements
as the amplified signal of a single entity, and to applications were small dispersions are
usually required, \eg, for magnetic recording media.

Complemental information on magnetism may be found in other reviews about magnetic
nanostructures\cite{bib-HIM98,bib-MAR03,bib-SKO03,bib-BOB04}. Finally, notice that this manuscript
does neither cover magnetic clusters fabricated by chemical routes\cite{bib-CRphys-Chaudret} and
their self-organization at surfaces, nor the use of pre-patterned substrates to align these
clusters along certain features\cite{bib-CHE04}.

\section{Magnetic order and thermal excitations in reduced dimension}
\label{sec-magneticOrder}

\begin{figure}[b]
  \begin{center}
  \includegraphics{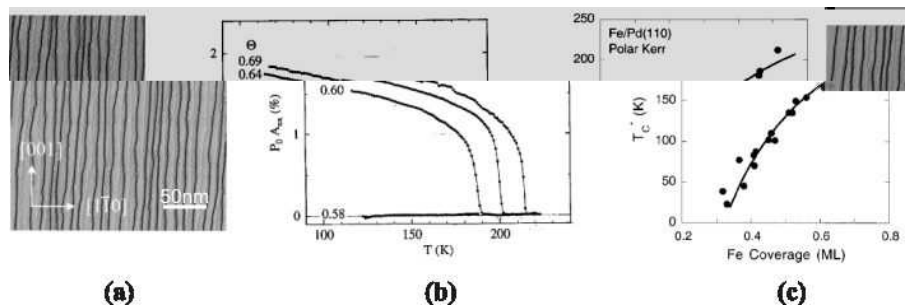}%
  \end{center}
  \caption{\label{fig-magneticOrder}(a)~differentiated STM image of \SO Fe/W(110) stripes with $\thickAL{0.5}$ coverage\cite{bib-HAU98b}
    (b)~in-plane remanent magnetization revealing a dramatic rise of \emph{apparent} $\Tc$~(in reality, $\Tb$, see text) upon percolation of \SA Fe/W(110) dots around $\thickAL{0.6}$ (c)~finite-size scaling for
    \SO Fe/Pd(110) stripes. The line is a fit with \eqnref{eq-finiteSizeScaling}\cite{bib-LI01}.}
\end{figure}

\subsection{Ferromagnetic ordering}
\label{subsec-magneticOrdering}

Continuous ultrathin films were used in the past as model systems to study magnetic ordering in
two dimensions~(2D). More recently \SA\ and \SO\ were used to fabricate systems with reduced
lateral dimensions, and thus study the cases of 1D~(stripes and wires) and 0D~(dots).

Spontaneous magnetic ordering in low dimension relies on the presence of \MAE. Already in two
dimensions long-range magnetic ordering occurs only for the Ising model\cite{bib-ONS44}. An $xy$
system is able to order in a finite two-dimensional system\cite{bib-BRA93b}, whereas a truly
isotropic Heisenberg system cannot establish long-range order at finite
temperatures\cite{bib-MER66}. The \MAE\ also influences the values of critical exponents. The rise
of spontaneous magnetization per unit volume $\Ms$ close below $\Tc$ follows the scaling law
$\Ms\propto(1-T/\Tc)^\beta$, with $\beta\approx0.125$ for 2D-Ising\cite{bib-ONS44} and
$\beta\approx0.23$ for 2D-$xy$\cite{bib-BRA93b}. These exponents have been measured in many
continuous ultrathin films and were found to agree with theoretical predictions of the Ising or
$xy$ models, depending on the type of \MAE, either uniaxial~(perpendicular or uniaxial in-plane)
or easy-in-plane\cite{bib-HIM98}.

The fabrication of 1D systems by \SO\ mostly relies on step-decoration of vicinal surfaces
in the step-flow growth regime with sub-atomic-layer~(\AL) amounts of
material\cite{bib-ELM89,bib-ELM94,bib-FIG95,bib-SHE97b,bib-DAL00,bib-GAM03b,bib-LI01}. The
evolution from 2D towards 1D was pioneered by Elmers \etal, who fabricated \SO\ Fe(110)
stripes on vicinal W(110)\cite{bib-ELM89,bib-ELM94} and evidenced the finite-size scaling
law\cite{bib-ALL70}

\begin{equation}
\label{eq-finiteSizeScaling} \Tc(n)/\Tc(\infty)\propto1-(n_0/n)^\lambda
\end{equation}

with $\lambda=1.03$ in excellent agreement with $\lambda=1$ predicted for an Ising system. $n$ is
the width of the stripes and $n_0$ the width for vanishing ordering at zero temperature,
tentatively extrapolated to four atoms although \eqnref{eq-finiteSizeScaling} is in principle not
applicable for small~$n$. $\lambda=1.2\pm0.3$ and $n_0=3$ was found for \SO\ Fe stripes on vicinal
Pd(110)\cite{bib-LI01}\bracketsubfigref{fig-magneticOrder}{c}. The temperature dependance of
magnetization was also carefully measured in the latter case, revealing enhanced magnetization
decay for decreasing width. In strictly one dimension even for an Ising system magnetic ordering
is not predicted at finite temperature, which is consistent with $n_0\simeq1$. However hysteresis
loops performed on \SO\ Co/Pt(997) chains of monoatomic width and high uniaxial \MAE, displayed
remanence and coercivity below \tempK{15}\cite{bib-GAM02}. This apparent contradiction is easily
lifted by noticing that ferromagnetic order can formally be defined only for systems of infinite
size and under thermodynamic equilibrium. In experiments the length of the wires and the duration
of the measurement are finite, so that single-domain states may occur when the correlation length
exceeds the physical length of the wire and no major thermal excitation occurs during the
measurement. Coercivity and remanence were also reported on flat finite-size
dots\cite{bib-RUS03,bib-WEI04b}. Provided that their lateral size is smaller than all
micromagnetic length scales and that the correlation length exceeds the dot's size, a nearly
uniform magnetization state is expected. This is a so-called near single-domain state that behaves
like a macrospin of moment $\mathcal{M}=\Ms\times V$ were $V$ is the volume of the system. Such
dots may be classified as 0D as there are fully described by the degree(s) of freedom of the
resulting macrospin only.

Apart from \MAE, dipolar interactions also affect ferromagnetic transition in low dimension.
Positive interactions can stabilize ferromagnetism because of their long range, despite being much
smaller in magnitude than exchange. The microscopic picture is the following. Exchange forces
establish large blocks of parallel spins, owing to their strength at low range. However in low
dimension these blocks may fluctuate on a large scale. The blocks can freeze thanks to the long
range of dipolar forces, despite their small strength, because what is to be compared with thermal
energy in the dipolar energy of large blocks, not of individual spins. This was confirmed, again
on an array of \SO\ Fe/W(110) stripes\cite{bib-HAU98b}\bracketsubfigref{fig-magneticOrder}{a}. In
contrast with the case of non-interacting stripes\cite{bib-ELM89} the signature of the dipolar
forces here was a sharp transition of $\Ms(T)$  around $Tc$ despite a significant dispersion of
stripes' width~(see the finite-size scaling above), and the absence of relaxation even just below
the freezing temperature. Because of the analogy with superparamagnetism, this effect was named
\emph{dipolar superferromagnetism}. This brings us naturally to the next paragraph.

\subsection{Superparamagnetism}
\label{subsec-superparamagnetism}

\begin{figure}[b]
  \begin{center}
  \includegraphics{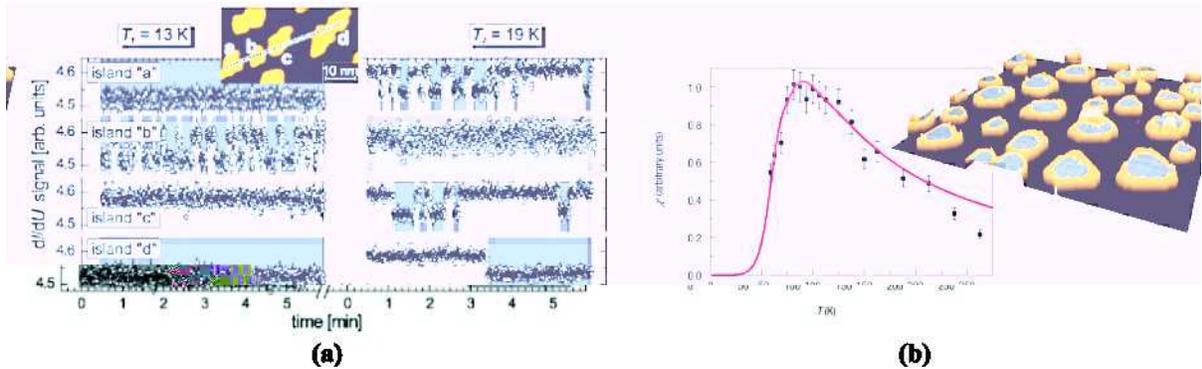}%
  \end{center}
  \caption{\label{fig-superpara}(a)~spontaneous magnetization switching
    of \SA Fe/Mo(110) dots~(see inset) close to $\Tb$, measured with
    Sp-STM\cite{bib-BOD04} (b)~$\Tb$ of \SA Co rings fabricated around Pt(111)
    dots~(inset) determined with susceptibility measurements\cite{bib-RUS03}}
\end{figure}

So far we have discussed experimental results in terms of a ferromagnetic transition, which is a
thermodynamic equilibrium concept. We have thus ignored kinetic effects, that we examine here, first theoretically, then experimentally.

Let us call ferromagnetic with Curie temperature $\Tc$ a system whose dimensions are all smaller
than the correlation length for $T<\Tc$. If a snapshot was experimentally feasible, it would
reveal essentially uniform magnetization, with a direction lying along an easy direction of
magnetization. At $T>\tempK0$ thermal fluctuations allow the system to explore its entire phase
space. Thus after a lapse of time $\tau$, $\boldsymbol{\mathcal{M}}$ would statistically have
overcome the energy barrier $E=KV$ associated with a hard  axis, and would have settled in another
easy axis direction. Using a simple Arrhenius law based on the Boltzmann probability of states
occupancy we have:

\begin{equation}
  \tau=\tau_0\exp{\beta KV}
\end{equation}

were $\beta=1/kT$ and $\tau_0\approx\unit[\scientific{}{-10}]{s}$\cite{bib-NEE49,bib-BRO63} is the
so-called attempt period. Let us assume that the system is observed for a duration~$\tau$. For
$T<\Tb$, with $\Tb=KV/[\kb\ln(\tau/\tau_0)]$ called the \emph{blocking temperature}, the system
has not switched over time $\tau$ and is said to be in a \emph{blocked state}. For $T>\Tb$ the
system switches spontaneously in the interval of time $\tau$. A magnetization measurement
performed over that time scale reveals zero apparent magnetization, although the magnitude of the
microscopic moment is still $\mathcal{M}$ at any time. This is the so-called
\emph{superparamagnetic} state. Analysis of remanent-less superparamagnetic magnetization loops
can be done with \emph{Brillouin-like} functions with $\mathcal{M}$ as an argument and allows one
to infer $\mathcal{M}$. From this value $V$ can generally be inferred because $\Ms$ is known.
Notice that owing to the high \MAE\ of most nanosized systems, in the range $\Tb<T<5\Tb$ the
function most fit to describe magnetization loops of magnetically-textured systems is closer to
the Brillouin~$1/2$ function than to the Langevin
function\cite{bib-CHA85,bib-FRU02b,bib-RUS03,bib-CRphys-Brune}. Using the latter leads to an
overestimation of $V$ by a factor~$3$. It is also well known that neglecting the size distribution
leads to an overestimation of the average $V$\cite{bib-RUS03,bib-CRphys-Brune}. As a second step
the measurement of $\Tb$ is sometimes used to deduce $K$. Let us emphasize that when the system's
size is larger than the wall width $\lambda$, magnetization reversal is \emph{not} coherent. In
such a case the volume involved in the determination of $\Tb$ is close to that of a nucleation
volume\cite{bib-BRO62,bib-MOO85}, of dimensions close to a wall width, not directly related to the
total volume of the system. This can lead to an important overestimation of $K$.

A first consequence of what is said above is that $\Tb<\Tc$, and what can be measured practically
in nearly all experiments is $\Tb$, not $\Tc$~(notice however that attempts have been made to
estimate $\Tc$ from the value of $\mathcal{M}$ inferred in the superparamagnetic regime,
associated with the coherence length and then compared to the volume of the
system\cite{bib-GAM02}). This is the case of the 1D system of Co/Pt(997) mono-atomic wires
reported above. The fact that $\Tb<\Tc$ was clearly illustrated in the case of in-plane magnetized
Fe/W(110) dots self-assembled in the sub-\AL\ range, and of lateral size $\thicknm{5-10}$. These
show an abrupt occurrence of spontaneous magnetization up to $\tempK{190}$ upon physical
percolation of the dots into a film for coverage $\Theta=\thickAL{0.6}$, whereas no spontaneous
magnetization was observed for $\Theta=\thickAL{0.58}$, down to $\tempK{115}$. This effect cannot
be explained by a finite-size scaling of the dots, and was attributed to
superparamagnetism\cite{bib-ELM94}\bracketsubfigref{fig-magneticOrder}{b}. This abrupt transition
was also observed for Co/Au(111) \SO dots ordered on a rectangular array, percolating from 0D to
1D and then from 1D to 2D\cite{bib-PAD99b}. Recently small Fe/Mo(110) dots with perpendicular
magnetization were magnetically imaged with spin-polarized STM~(Sp-STM)~(for the technique see
\cite{bib-BOD03}), revealing a blocked state below approximately
$\tempK{20}$\cite{bib-BOD04}\bracketsubfigref{fig-superpara}{a}. The observation of telegraph
noise between two states of high magnetization confirms that the macroscopic vanishing spontaneous
moment is related to $\Tb$, not to $\Tc$. Notice that thermal activation of single nanosized
objects was reported before and analyzed quantitatively with, \eg, micro-SQUID, a highly precise
technique\cite{bib-WER01}. Direct information about the measured nanostructure however was
lacking, which can now be brought with nanostructures grown at surfaces. The correlation between
structure and magnetic properties is a traditional cornerstone of material improvement, and the
same statement is expected to hold for nanomagnetism. An advantage of \SA\ combined with a
high-resolution direct imaging technique is that subtle changes in $\Tb$ can be associated with
the shape of dots, more compact dots displaying a higher $\Tb$, explained by a non-uniform
magnetization reversal process. Systems with a strongly non-compact shape were measured, like Co
rings fabricated by step-decoration of Pt(111) dots\bracketsubfigref{fig-superpara}{b}.

\section{Orbital moment and anisotropy, from bulk to single atoms}
\label{sec-anisotropy}

In 1954 N{\'e}el predicted that symmetry breaking at the interfaces of magnetic thin films would
induce an extra contribution to the \MAE, named interface magnetic anisotropy, or N{\'e}el anisotropy.
It yields a $1/t$ dependance of the total \MAE\ density of a film with
thickness~$t$\cite{bib-NEE54}. The experimental confirmation of this law was given in
1968\cite{bib-GRA68}, with the foreseen possibility to overcome dipolar energy and yield
perpendicular magnetization\cite{bib-CRphys-neelAnisotropy}. \MAE\ was then predicted to scale
with the anisotropy of the orbital momentum\cite{bib-BRU89b}, which was checked experimentally on
perpendicularly magnetized Au/Co/Au(111)\cite{bib-WEL95}. See Ref.\cite{bib-STO99} for a review.

\SA\ and \SO\ opened the possibility to study this phenomenon quantitatively in even lower
dimensions, \ie, the anisotropy associated with atomic edges~(1D) and kinks~(0D). This issue is of
direct interest for applied nanomagnetism, as in ever smaller grains the main source of \MAE\ will
arise from surfaces, edges and kinks. A widely used technique is X-ray magnetic circular dichroism
for its ability to separate spin from orbital momentum through the use of sum
rules\cite{bib-STO99}, and its high sensitivity, well below one atomic
layer\cite{bib-GAM02b,bib-GAM03}. Pioneering work was done on \SO\ Co/Au(111)
dots\cite{bib-DUR99,bib-KOI01}. Spin and orbital momentum were recorded as a function of the dot's
diameter~$L$. The $1/L$ dependance normalized per atom was then interpreted as an edge
contribution. The spin contribution was reported to be mostly unaffected at edges in all
publications. The orbital momentum was found to be non-affected at edges by Koide \etal, whereas
an increase of orbital momentum was reported by D{\"u}rr \etal In this publication if one uses the
relevant function for fitting superparamagnetic curves to assess the volume of dots, \ie,
Brillouin $1/2$ instead of Langevin used in the publication~(see \secref{sec-micromagnetism}), we
can estimate the ratio of edge over surface atoms for each measurement. Then we deduce that edge
atoms bear an extra orbital moment of $\unit[0.5\pm0.2]{\muB}$ as compared to bulk Co. A rise of
orbital momentum of $\unit[0.5]{\muB}$ was also determined in \SO\ Fe/Au(111)
dots\cite{bib-OHR01}, in quantitative agreement with the value of enhanced magnetization at steps
on thin films\cite{bib-ALB92,bib-GRA97}. However in this case the interpretation is more difficult
owing to the transition of Fe from a high-spin to a low-spin phase as a function of dot's size.

The real breakthrough came from the study of Co/Pt systems. First, the \SO of Co was optimized on
the vicinal Pt(997) surface, yielding stripes of width seven atoms for \unit[1]{\AL coverage},
down to mono-atomic width, \ie, wires\cite{bib-DAL00}. Orbital momentum, its anisotropy, and the
\MAE\ could be measured as a function of the stripe width. The phenomenological contribution of
edge atoms was then extracted\cite{bib-GAM02,bib-GAM03b}. An oscillatory behavior of \MAE\ was
also found\cite{bib-GAM04}, an effect similar to oscillatory behaviors well known in 2D. Similar
studies were performed on small flat dots of Co/Pt(111) from one to some tens of atoms. The
control of the size of the dots was achieved by first depositing at low temperature~(\tempK{15}),
where the motion of adatoms is hindered and thus only single atoms are found on the surface,
followed by careful step-like annealings to increase the mean dot's size by diffusion-limited
aggregation\cite{bib-GAM03,bib-CRphys-Brune}~(Notice that single atoms were first probed by XMCD
prior to these experiments, for 3d atoms on alkali surfaces\cite{bib-GAM02b}). The outcome of
various experiments from bulk to single atoms is summarized in \tabref{tab-coOrbitalMoment}. The
orbital momentum, essentially quenched in the bulk, rises progressively above $\unit[1]{\muB}$ in
single atoms, along with the \MAE. It is found that, from the point of view of orbital momentum
and \MAE, a cluster of two atoms~(a bi-atomic wire, resp.) behaves closer to a wire~(a mono-atomic
slab, resp.) than to a single atom~(a mono-atomic wire, resp.)\brackettabref{tab-coOrbitalMoment}.
This reminds us that the concept of dimensionality depends on the magnetic property studied: two
atoms is closer to a wire when orbital momentum is concerned, but is closer to a single atom when
magnetic ordering is concerned, see \secref{subsec-magneticOrdering}.

{
\begin{table}[t]
  \centering
  \caption{Orbital momentum and magnetic anisotropy energy~(MAE) of Co atoms on Pt as a function of coordination
    (after \cite{bib-GAM02,bib-GAM03}).}
  \label{tab-coOrbitalMoment}
  \newlength{\widthItem}
  \def\textItem{~~Orbital momentum~~}
  \settowidth{\widthItem}{\textItem}
  \newlength{\widthItemm}
  \def\textItem{~~bi-atomic~~}
  \settowidth{\widthItemm}{\textItem}
  \newlength{\widthItemmm}
  \def\textItem{~~mono-atomic~~}
  \settowidth{\widthItemmm}{\textItem}
  \begin{tabular}{p{\widthItem}ccp{\widthItemm}p{\widthItemmm}cc}
    \hline\hline    &  ~~bulk~~  & ~~mono-layer~~ & \centering ~~bi-atomic wire~~  & \centering ~~mono-atomic wire~~  & ~~two atoms~~ &
    ~~single atom~~\\
    \hline Orbital momentum ($\muB/\mathrm{at}$) & 0.14 & 0.31 & \centering 0.37 & \centering 0.68 & 0.78 & 1.13 \\
    MAE ($\mathrm{meV}/\mathrm{at}$) & 0.04 & 0.14 & \centering 0.34 & \centering 2.0 & 3.4 & 9.2
    \\\hline
  \end{tabular}
\end{table}
}

Model \SO\ and \SA\ systems like the Co/Pt structures presented above, are desirable to extract
quantitatively fundamental properties like \MAE. This knowledge being established, the question
arises to what extent it can be used to tailor the \MAE\ of more versatile systems suitable for
applications. It is sometimes argued that the increase of \MAE\ in low dimensions will help to
overcome superparamagnetism. The evaluation of the energy of nucleation
volumes\bracketsecref{subsec-superparamagnetism} allows one to draw general trends, revealing that
this is not the case. Let us assume that the \MAE\ in any system is dominated by the contribution
of lowest dimensionality, \ie, surface $\Ks$ for films and edges $\Ke$ for stripes. One finds that
for continuous films of thickness $t$~($D=2$) $\Tb$ is essentially independent of $\Ks$ and
$\Tb\propto t$; Concerning $D=1$, for wires of more or less round section $L\times L$, $\Tb\propto
L^{3/2}$, while for flat wires of width $L$, $\Tb\propto L^{3/4}$. Finally for $D=0$, $\Tb\propto
L^2$ for compact clusters and $\Tb\propto L$ for flat dots. Thus, in all cases a \emph{decrease}
of $\Tb$ is expected upon decrease of the system's size. The only hope resides in increasing the
number of interfaces in a system of fixed size. Engineering \SA\ to produce lateral superlattices
via sequential deposition is a promising way. It has been demonstrated for 1D single rings for
magnetic materials\cite{bib-RUS03,bib-CRphys-Brune}~(more generally, multiple concentric rings and
lateral stripes of Ge/Si\cite{bib-KAW04} were reported). This concept applies also to the case of
spontaneously ordered lateral multilayers of FeAg and CoAg\cite{bib-TOB98,bib-TOB00}, see
\secref{sec-material}. Ultimately no strict borderline exists between this approach and the
fabrication of high-anisotropy ordered alloys like FePt\cite{bib-SUN00} or CoPt\cite{bib-ALB01}.
Tailoring the anisotropy with arrays of interfacial dislocations is another approach, as reported
for Fe/W(001) nanostructures\cite{bib-WUL00,bib-WUL03}.

\section{Model systems for micromagnetism}
\label{sec-micromagnetism}

\begin{figure}[b]
  \begin{center}
  \includegraphics{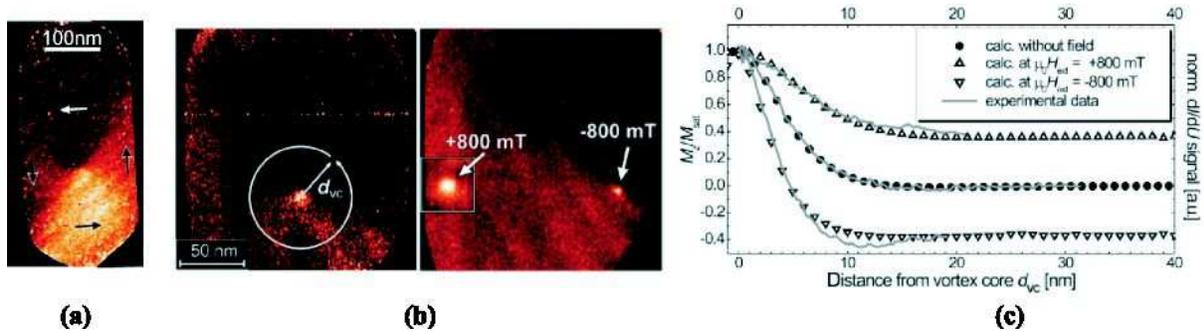}%
  \end{center}
  \caption{\label{fig-micromagnetism}Micromagnetism with Sp-STM using an Fe/W(110) dot\cite{bib-WAC02}.
    (a)~overview of the flux-closure domain state, in-plane surface magnetization (b)~out-of-plane component for zero-field~(left),
    and in-perpendicular-field~(right, main for negative field, inset for positive field).
    (c)~surface profile of the vortex core.}
\end{figure}

Micromagnetism is the field of study of magnetic domains and domain walls, both statically and
under magnetization reversal or magnetic excitations, that typically spans in the range
$\unit[50]{nm}-\thickmicron{10}$. Lithography is then the most relevant fabrication technique
because of its resolution and versatility\cite{bib-MAR03,bib-HUB98b}. Nevertheless UHV-fabricated
nanostructures have niche applications, where they are better suited: 1-~specific techniques can
be applied that require a UHV-compatible surface. This is the case of Sp-STM, that allows both the
highest available magnetic resolution, down to the atomic level\cite{bib-BOD03}, and the
application of an external field of arbitrary value in any direction of space, which can mostly
not be done with high-resolution microscopies based on electrons~(SEMPA or spin-SEM, Lorentz,
X-PEEM and SPLEEM) 2-~surfaces and edges can be fabricated avoiding defects arising during
lithography and/or etching~(amorphisation, oxidation, etching or resist-related loss of
resolution), thus yielding high-quality nanostructures for model investigations 3-~UHV fabrication
may be more reliable than lithography, \eg, for nanometer-sized features\cite{bib-PIE04} or, on
the reverse, objects with a high vertical aspect ratio\cite{bib-FRU03c}. All these aspects are
illustrated below.

The core of magnetic vortices could be imaged by Sp-STM in flux-closure
states\cite{bib-WAC02}\bracketfigref{fig-micromagnetism}. Magnetic vortices had previously been
detected by magnetic force microscopy in dots made by lithography\cite{bib-SHI00b}. However, UHV
measurements with the high-resolution Sp-STM technique yielded the true width of the vortex core
at the surface, $\thicknm{9\pm1}$ for Fe(\thicknm{8})/W(110), in good agreement with micromagnetic
predictions\cite{bib-HUB98b}. The size of the core was found to expand~(resp. shrink) upon
application of an external field parallel~(resp. antiparallel) to the magnetization in the core,
again in good quantitative agreement with micromagnetic predictions. Another topic is geometrical
constrictions. Bruno predicted that domain walls may be compressed in geometrical constrictions,
because the increase of exchange energy can be overcompensated by a decrease of the length of the
wall, thus of its total energy\cite{bib-BRU99b}. The shrinking of domain walls in constrictions
was confirmed in \SO double-layers~(DLs) Fe/W(110) stripes, \eg, from $\thicknm{6}$ in smooth
stripes to $\thicknm{2}$ for a constriction $\sim\thicknm{1}$ wide and long\cite{bib-PIE00}. Going
to ever smaller scale, as conventional micromagnetism is a continuum theory, it breaks down at the
atomic level. Discrete models may be substituted, \eg, to describe extremely narrow domain walls
in hard magnetic materials\cite{bib-HIL73}. Such narrow domain walls could be imaged with Sp-STM
on Fe(\thickAL{1})/W(110), revealing a width of $\thicknm{0.6}$ using the asymptote of the cosine
angle, a value that may still be limited by the spatial resolution of the technique. In such high
anisotropy materials the wall profile depends solely on exchange~$A$ and \MAE~$K$. Thus, based on
values of $K$ measured by techniques like ferromagnetic resonance or torque-oscillatory
magnetometer that can be applied down to the monolayer\cite{bib-GRA93}, values of the exchange
could be extracted. On a larger scale, around $\thicknm{100}$, geometrically-constrained domain
walls were studied in cross-paths of \SA\ Fe/W(001) wires using SEMPA, revealing the arrangements
for the four possible topological incoming magnetic fluxes from the four arms\cite{bib-WUL03}.

The above-mentioned reports concern very small length scales. Another direction of research in
micromagnetism consists in studying ever larger systems, to try to bridge the gap between
nanostructures now understood quantitatively, and macroscopic materials still described
phenomenologically. The use of \SA\ is justified as larger systems become increasingly complex, so
that it is essential to study model nanostructures to avoid an extra interplay of extrinsic
effects related to defects. Studies at zero field concern the direct observation of the transition
from the single-domain to vortex-state\cite{bib-JUB04} in \SA Fe/W(001) dots with a more or less
square shape\cite{bib-YAM03}, more generally the evolution from single-domain to a variety of
flux-closure domain states like vortex, Landau and diamond\cite{bib-HUB98b} in \SA\ elongated
Fe/W(110) dots\cite{bib-FRU01b,bib-FRU03c,bib-BOD04b}. Thermally-activated switching between
metastable states was investigated in Co/Ru(0001) \SA\ dots, namely single-domain and the
so-called V-state\cite{bib-DIN04}. A further step consists of the quantitative understanding of
magnetization reversal in such multi-domain states. Nucleation and annihilation fields of magnetic
vortices and walls were measured with a micro-SQUID in a single dot $\thicknm{30}$-thick as a
function of the in-plane direction of the external field. Discontinuities were evidenced as a
function of angle. Such jumps are sometimes thought to result from defects. Here, they could be
ascribed, with the help of simulations, to intrinsic bifurcations related to the direction of the
external field with respect to edges\cite{bib-FRU04c}. Very few experiments are available for
thicker--thus more bulk-like--dots\cite{bib-HEH96} whereas such systems can now be tackled with
simulations\cite{bib-RAV98b}. Such studies are now in progress, revealing features that cannot be
understood with 2D micromagnetics only\cite{bib-FRU05}.

\section{Thick self-organized systems: from surfaces to materials}
\label{sec-thick}

\begin{figure}[b]
  \begin{center}
  \includegraphics{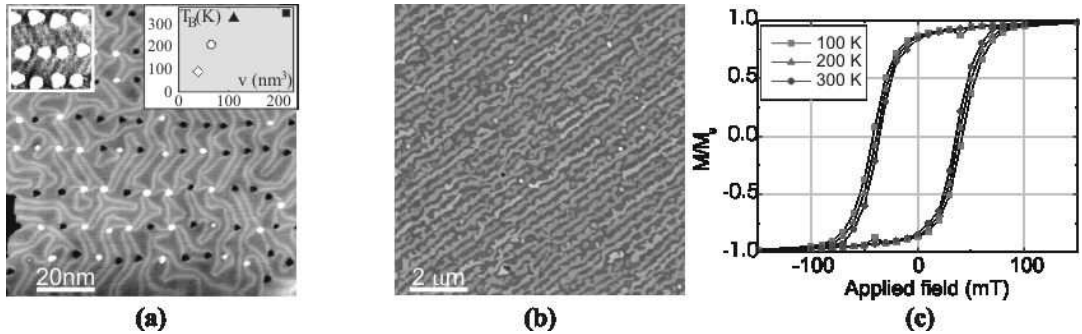}%
  \end{center}
  \caption{\label{fig-thick}(a)~STM image of a \SO\ Co/Au(111) system during the fabrication of pillars. Insets:
  Conventional \SO Co/Au(111) dots in the sub-AL range~(left) and blocking
  temperature of pillars as a function of their volume\cite{bib-FRU02b}~(right) (b)~\SO Fe/W/Mo(110) stripes,
  $\thicknm{4.3}$-thick\cite{bib-FRU04} (c)~magnetization loops of Fe/W(110), $\thicknm{5.5}$-thick, along in-plane [001].}
\end{figure}

It is still unclear whether epitaxial \SA\ or \SO\ of magnetic systems will bring any new useful
functionality. Above all, there exists at the moment a fundamental obstacle against their use in
devices. These systems have small lateral dimensions and are generally one or two \ALs high only,
so that they have an extremely small volume. Thus, they do not provide enough material for
applications, and most important, they are all superparamagnetic at room
temperature\cite{bib-SHE97b,bib-TAK97,bib-PAD99b,bib-GAM02,bib-GAM03}, see
\secref{sec-anisotropy}. We have discussed that the rise of \MAE\ in low dimension is overbalanced
by the decrease of volume, so that the issue becomes ever more acute for smaller nanostructures as
the anisotropy barrier $KV$ shrinks. At first sight $V$ cannot be increased much, as for
conventional deposition processes upon increasing the amount of material deposited percolation
into a continuous film occurs\cite{bib-ELM94,bib-PAD99b,bib-GAM02}. The key may lie in engineered
growth processes that were demonstrated to yield \SO\ nanostructures much thicker than atomic
layers while avoiding percolation, thus significantly increasing $V$ without compromise on the
lateral density, see below.

The first process was demonstrated for \SO Co/Au(111) dots. Inspired by the vertical stacking of
multilayers of quantum dots\cite{bib-CRphys-Springholz} sequential deposition of \AL-fractions of
Au and Co was performed. When Au just fills up the empty space between the Co dots, the dots from
the next layer grow atop the existing dots, thereby increasing their height by one~\AL. This
process is driven by immiscibility and a large lattice mismatch. Multilayers thus yielded pillars
of height up to 8nm and diameter in the range
$\thicknm{3-5}$\cite{bib-FRU99d,bib-FRU00}\bracketsubfigref{fig-thick}{a}. Magnetization is
essentially perpendicular, $\Tb$ increases monotonically with $V$ and could be brought up to
$\tempK{350}$\cite{bib-FRU02b}.

Another process makes use of strain fields arising in the vicinity of atomic steps. Upon
deposition of Fe on vicinal cc(110)~(cc=Mo,W) at the temperature of layer-by-layer
deposition, a smooth film is formed except above the buried atomic steps of cc were
trenches are formed. Then, upon annealing unwetting of the substrate occurs in registry
with the trenches, yielding an ordered array of stripes with heights up to
$\thicknm{5}$\cite{bib-FRU04}\bracketsubfigref{fig-thick}{b}. Magnetically soft or
hard\bracketsubfigref{fig-thick}{c} stripes were obtained by tuning the interface
anisotropy. Thus the two main functional properties of ferromagnets could be demonstrated
at room temperature, first coercivity-remanence, second the ability to break down into
(stable) domains, whereas conventional Fe/W(110) stripes have
$\Tc=\tempK{179}$\cite{bib-HAU98b}. Reports of trenches atop buried steps on other
systems\cite{bib-CHE01,bib-FRU02d} suggest that this process might be quite general.

\section{Self-organization for material design}
\label{sec-material}

\begin{figure}[b]
  \begin{center}
  \includegraphics{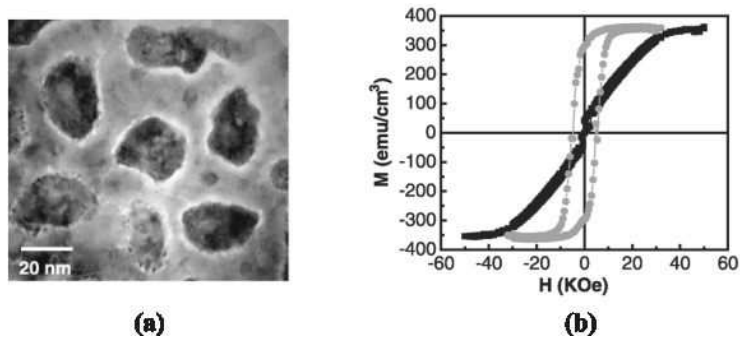}%
  \end{center}
  \caption{\label{fig-materials}\SO ferrimagnetic $\mathrm{CoFe}_2\mathrm{O}_4$ columns in a
   $\mathrm{BaTiO}_3$ ferroelectric matrix. (a)~plane view (b)~magnetization loops performed at $\tempK{300}$ for
   perpendicular~(shaded) and in-plane~(dark) field\cite{bib-ZHE04}.}
\end{figure}

Systems reviewed in sections \ref{sec-magneticOrder} through \ref{sec-micromagnetism} were used to
study fundamental properties of nanosized objects. Concerning potential applications, \SO\ and
\SA\ systems are more liable to be used as a whole, to achieve a material with specific
properties, rather than for the direct use of each nanostructure independently. For instance, the
anisotropy of continuous films can be tailored by deposition on patterned surfaces like
step-bunched\cite{bib-ENC99} and facetted\cite{bib-ROU02b} surfaces, on surfaces grooved in
arbitrary directions after grazing-incidence ion sputtering\cite{bib-MOR03}, by shadow deposition
on facetted surfaces\cite{bib-SUG97,bib-TEI02,bib-WEI04}.

 Several types of systems with tailored magnetic properties can be fabricated by \SO.
Concerning \SO\ from the deposit, arrays of parallel nanometer-sized Fe-Ag stripes and Co-Ag
stripes are formed by co-deposition on Mo(110)\cite{bib-TOB98}. These lateral superlattices
display a high in-plane \MAE\ along the wires, and magneto-transport with cpp geometry while the
current flows in-the-plane\cite{bib-TOB00}. The highest achieved $\Tb$ in these systems is
currently $\tempK{220}$, with an in-plane anisotropy field around
$\unit[0.5]T$\cite{bib-CRphys-BOR}. The growth of an FeIr/Ir(001) lateral superlattice with
$\sim\thicknm{1}$-period was also reported\cite{bib-KLE04}. We have already noticed that there
exists no strict borderline between nanoscale superlattices fabricated by \SO and layered ordered
alloys of high \MAE\ like $\mathrm{L}_10$ phases\cite{bib-SUN00}. Concerning these alloys, \SA\
can be used to lower the ordering temperature owing to enhanced adatom mobility along the facets
of dots\cite{bib-ALB01}. This initial stage of nucleation can be used to fabricate at moderate
temperature and upon percolation a grainy continuous film, partially ordered and thus with
perpendicular anisotropy, found below $\thicknm{8}$\cite{bib-ALB02}.

In \secref{sec-thick} I have not included columnar growth because no clear borderline exists
between \SA\ and continuous films like CoCr-based columnar recording media, where \SA\ can indeed
be used to create a proper microstructure to adjust extrinsic properties like coercivity. Recent
developments is columnar growth for oxides, \eg\,
$\mathrm{La}_{0.8}\mathrm{Sr}_{0.2}\mathrm{MnO}_3/\mathrm{LaAlO}_3$\cite{bib-JIA02} and
$\mathrm{CoFe}_2\mathrm{O}_4$ ferrimagnetic columns with high perpendicular anisotropy due to the
strain imposed by a $\mathrm{BaTiO}_3$ ferroelectric
matrix\cite{bib-ZHE04}\bracketfigref{fig-materials}. One order of magnitude higher density could
be readily achieved for hard-disk media with present technology if monodisperse media were
available with no ordering defects, motivating demonstrations for such patterns of dots, like for
Co/Au(788)\cite{bib-REP02b,bib-WEI04b}. An interesting area to follow in the future is the
combination of prestructuring, \eg, using lithography, and self-assembly, an already ripe domain
for semiconductors\cite{bib-CRphys-Eymery} but emerging for metals\cite{bib-YU01,bib-CHE04}. This
would combine the versatility of lithography with the model features of UHV-deposited
nanostructures.

\section{Conclusion}

This manuscript proposed a short overview of the use of self-organization~(\SO) and
self-assembly~(\SA) for magnetic purposes. The studies performed fall into two categories. In the
first category \SO\ and \SA\ are used to measured and analyze quantitatively low-dimensional
magnetic phenomena. Beyond fundamental knowledge, these phenomena occur in systems of interest for
application. However, they cannot be studied directly in these because of their extreme
complexity, in terms of distributions, defects and microstructure. Thus, the idea is to get
fundamental knowledge from surfaces, to be used to further tailor the properties of functional
materials. The advantages of \SO\ and \SA\ over lithography are twofold. First, the high surface
quality and cleanness allows intrinsic phenomena to be recorded, and also demanding techniques
such as spin-polarized scanning tunneling microscopy can be applied. Second, the achievable size,
down to the atomic level, lies much below that of any existing lithography technique. Issues that
were investigated and reported here are magnetic ordering in 1D or 0D, spin and orbital momentum
at edges and kinks, superparamagnetism in 1D or 0D. In the second category studies aim at bridging
the gap between surfaces and materials, \ie, at showing that \SO\ and \SA\ systems might be useful
directly in devices. These include the fight against superparamagnetism, mainly by demonstrating
processes for vertical growth of \SA and \SO nanostructures, and the tailoring of magnetic
properties like anisotropy and the distribution of properties.



\section*{Acknowledgements}
I am grateful to W. Wulfhekel and F. Scheurer for a critical reading of the manuscript.

\bibliographystyle{report-nodescription}
\bibliography{fruche4,fruchart}

\begin{thebibliography}{100}
\bibitem{bib-CRphys-SOSA}
\texttt{In the case of the spontaneous formation of nanostructures upon
  deposition at a surface I call \SA\ the case were the nanostructures display
  no or only short-range positional order, and \SO\ the case were the
  nanostructures display a positional order with at least a middle-range order,
  typically over at least 5-10 items. Notice that this vocabulary is not
  universaly used in the literature, and may also vary from one community to
  another. We will abusively use the acronym \SO~(resp. \SA) to stand for
  different expressions like self-organization, self-organize, self-organized.}
\end{thebibliography}







\end{document}